
\magnification =1200
\openup 2\jot
\hsize=15.4 truecm\vsize=23.4 truecm
\pageno =0
\parindent=0 cm

\input epsf

\font\list=cmcsc10
\font\cat=cmr7


\newcount\refno
\refno=0
\def\nref#1\par{\advance\refno by1\item{[\the\refno]~}#1}

\def\book#1[[#2]]{{\it#1\/} (#2).}

\def\annph#1 #2 #3.{{\it Ann.\ Phys.\ (N.\thinspace Y.) \bf#1}, #2 (#3).}
\def\cmp#1 #2 #3.{{\it Commun.\ Math.\ Phys.\ \bf#1}, #2 (#3).}
\def\cqg#1 #2 #3.{{\it Class.\ Quan.\ Grav.\ \bf#1}, #2 (#3).}
\def\mpla#1 #2 #3.{{\it Mod.\ Phys.\ Lett.\ \rm A\bf#1}, #2 (#3).}
\def\ncim#1 #2 #3.{{\it Nuovo Cim.\ \bf#1\/} #2 (#3).}
\def\npb#1 #2 #3.{{\it Nucl.\ Phys.\ \rm B\bf#1}, #2 (#3).}
\def\plb#1 #2 #3.{{\it Phys.\ Lett.\ \rm B\bf#1}, #2 (#3).}
\def\pla#1 #2 #3.{{\it Phys.\ Lett.\ \rm A\bf#1}, #2 (#3).}
\def\prd#1 #2 #3.{{\it Phys.\ Rev.\ \rm D\bf#1}, #2 (#3).}
\def\prl#1 #2 #3.{{\it Phys.\ Rev.\ Lett.\ \bf#1}, #2 (#3).}

\def\bo{ { \sqcup\llap{ $\sqcap$} } }
\overfullrule=0pt       
\def\L{$\cal L$}

\def\real{I\negthinspace R}

\def\half{\textstyle{1\over2}}

\def\spose#1{\hbox to 0pt{#1\hss}}
\def\lta{\mathrel{\spose{\lower 3pt\hbox{$\mathchar"218$}}
     \raise 2.0pt\hbox{$\mathchar"13C$}}}

\hbox{ }
\rightline {hep-th/9410095}
\rightline {DAMTP/R-94/41}
\vskip 5truemm

\centerline{\bf CURVATURE CORRECTIONS TO DYNAMICS OF DOMAIN WALLS}
\vskip 5mm

\centerline{ {\bf Brandon Carter}\footnote{$^\spadesuit$}{\sl Email:
carter@mesiob.obspm.fr } } \vskip 2mm

\centerline{\it Newton Institute for the Mathematical Sciences,
University of Cambridge}
\centerline{\it 20 Clarkson Road, Cambridge, U.K. CB3 0EH}
\centerline{\it and}
\centerline{ \it D.A.R.C., Observatoire de Paris, 92 Meudon, France. }

\vskip 3mm
\centerline{\bf and}
\vskip 3mm

\centerline{{\bf Ruth Gregory}\footnote{$^\clubsuit$}{\sl Email:
rg10012@amtp.cam.ac.uk}}
 \vskip 2mm

\centerline{ \it D.A.M.T.P. University of Cambridge, 
 Silver Street, Cambridge, CB3 9EW, U.K.}

\vskip 1.0cm
\centerline{\list ABSTRACT}
\vskip 3mm

{ \leftskip 5truemm \rightskip 5truemm

\openup -2\jot

The most usual procedure for deriving curvature corrections to effective
actions for topological defects is subjected to a critical reappraisal. A
logically unjustified step (leading to  overdetermination) is identified and
rectified, taking the standard domain wall case as an illustrative example.
Using the appropriately corrected procedure, we obtain a new exact
(analytic) expression for the corresponding effective action contribution of
quadratic order in the wall width, in terms of the intrinsic Ricci scalar
$R$ and the extrinsic curvature scalar $K$. The result is proportional to
$cK^2-R$ with the  coefficient given by $c\simeq 2$. The resulting form of
the ensuing dynamical equations is obtained in terms of the second
fundamental form and the Dalembertian of its trace, K. It is argued that
this does not invalidate the physical conclusions obtained from
the ``zero rigidity" ansatz $c=0$ used in previous work.

\openup 2\jot

}


\vfill\eject
\footline={\hss\tenrm\folio\hss}

\parindent=1cm
\noindent{\bf 1. Introduction: the importance of defect dynamics.}
\medskip

Topological and other vacuum defects are of interest and importance in many
areas of physics today. In high energy physics they generically
occur during a symmetry breaking process where different parts of a medium
choose different minimal energy configurations, or vacua, and the
non-compatibility of these different vacua forces a sheet, line, or point of
energy in which the vacua meet at a {\it defect}, where the relevant
vacuum order parameter becomes indeterminate. The phrase
topological defect is used to embody the idea that it is the topology of the
vacuum that simultaneously allows formation, and prevents dissipation, of
these objects -- but a defect need not be topological. Many instances are
known where a defect may be stable dynamically (i.e.~classically, due to
energy considerations) but not topologically: for example semilocal
defects{[1]} fall into this category. A defect can even be ``topological"
and unstable, as in the case of textures{[2]}, but nonetheless of physical
importance.

In cosmology, there are two main concerns when considering defects. One is
their gravity, and the other their dynamics. Any theory concerning large
scale transport of matter (such as in galaxy formation) must be able to
allow for, constrain, or even rule out, the presence of strongly
self-gravitating objects. But the primary concern still is dynamics. There
may be defects (such as low mass cosmic strings{[3]}) that have little
impact gravitationally when in straight static configurations, but that
become gravitationally important when strongly curved, crinkled, or compact
(as in the case of string loops). Questions of dynamics may also have an
essential influence on decay rates. It is therefore worthwhile to study the
purely dynamical aspects as a subject in its own right, leaving
gravitational aspects to be included in subsequent investigations. It is
this strategy that will be followed in the present analysis, whose scope
will be limited to defects in a flat Minkowski background in the interests
of conceptual clarity and mathematical simplicity.

Attempts to derive effective actions or equations of motion for topological
defects have commonly focussed on the strong coupling limit, meaning that of
large values of the coupling coefficient, $\lambda$ of the relevant Higgs
field. In this limit, the defect becomes infinitesimally thin and
effectively decouples from the other (infinitely massive) particles in the
field theory. The study of the effective motion of topological defects has
been extended{[4 -- 12]} away from the limit $\lambda\rightarrow 0$ to
cases for which the thickness is small but not exactly, zero. The resultant
effective action generically contains a ``zero-thickness'' term proportional
to the area of the defect, and extrinsic curvature terms which appear at
quadratic order in the thickness. It is the controversy about the way to
evaluate these second order terms that has prompted the present work.

While the earliest investigations {[4 -- 6]} agreed in predicting that
such extrinsic curvature terms should definitely exist, they failed to reach
consensus, not only about their amplitudes and their completeness (meaning
whether or not other ``twist" terms of the same order were needed as well)
but even about whether their signs corresponded to ``rigidity" or
``antirigidity". The confusion became worse{[7]} after the publication of many
subsequent studies{[8 -- 12]} predicting or assuming ``zero rigidity",
meaning the absence of any quadratic order corrections except for the term
proportional to the {\it intrinsic} curvature term, $R$ (which in the
case of a string is a pure divergence having no effect on the motion).

The present work makes a fresh start on the basis of a critical examination
of the procedure used in the preceding work {[4 -- 12]} in the simplest
case, namely that of a domain wall (for which the question of a ``twist"
contribution does not even arise). We adopt a ``classical" approach, that is
appropriate to a  strong coupling limit, $\lambda\rightarrow\infty$. However
the validity of our analysis is by no means  restricted to this limit, but
extends to moderate and even small values of $\lambda$. Our results are
applicable quite generally  to any limit in which the  wall curvature scale,
$L$ say, is large compared to the wall width, $\ell$ say, even when the
latter is not infinitesimal (though the method will not describe the
interaction of the wall with the underlying scalar field).

We find that the approach used in the original investigations{[4 -- 6]} was
essentially sound, their discrepancies being mainly due to the difficulty of
being sure that no terms were overlooked. However, while justified in having
doubted the detailed conclusions of these pioneering investigations, the
subsequent papers{[8 -- 10]} strayed from strict logic in imposing an
unduly restrictive simplification ansatz.

The present work corrects this step, providing a new evaluation of the
second order curvature contribution to the off-shell action, in the case of
a simple domain wall. The advantage of considering the domain wall becomes
apparent at this stage, for we are able to perform all operations
analytically, obtaining exact values for all the parameters in the second
order effective action.  It is found that the internal mechanics of the wall
is characterised by a well defined and {\it strictly negative} ``rigidity"
coefficient. This does not invalidate the use  in the previous work{[8 --
12]} of the corresponding ``zero rigidity" model as a permissible (though not
obligatory) second order approximation, because the effect on the dynamical
equations of the rigidity term in question is of higher order.

\bigskip\noindent
{\bf 2. The scalar field model.}
\medskip

The simplest relativistic domain wall model in common use is based on
a bosonic field theory consisting of a real scalar $\Phi$ whose self
interaction is governed by the Lagrangian
$$
{\hat{\cal L}}=-{\half}\big(\nabla_{\!\mu}\Phi\big)\nabla^\mu\Phi
-\lambda\big(\Phi^2-\eta^2\big)^2
\eqno(1)
$$
for positive constants $\eta$ and $\lambda$, in a $D+1$
dimensional background, with coordinates $x^\mu$ ($\mu=$ 0,1, ...,$D$), and
Lorentz signature (--, +, ... , +) spacetime metric $g_{\mu\nu}$.
In the present work this metric is postulated to be {\it flat} (which means
that gravitational effects are neglected). The Lagrangian (1) gives the well
known field equation
$$
\nabla_{\!\mu}\nabla^\mu\Phi-4\lambda\Phi\big(\Phi^2-\eta^2\big)=0
\eqno(2)
$$
which has two distinct homogeneous ``vacuum" solutions, $\Phi=\pm\eta$.
Positive and negative domains, as characterised respectively by $\Phi>0$ and
$\Phi<0$, are separated by domain walls that are identified as
hypersurfaces, with internal coordinates $\sigma^i$, ($i=$ 0, ..., {\sevenrm
D}-1), on which $\Phi=0$.

The simplest domain wall solution is given by the static plane wall ansatz
expressible in terms of Minkowski background coordinates $x^\mu$ by
$$
x^i=\sigma^i\ ,\hskip 1.6 cm  x^{_{\rm D}}=0\ ,
\hskip 1.6 cm \nabla_{\!i}\phi=0\ .
\eqno(3)
$$
Writing $z=x^{_{\rm D}}$ for
the last coordinate (the only one that is not ignorable), the field
equation reduces in this case to
$$
{d^2\Phi\over dz^2}-4\lambda\Phi\big(\Phi^2-\eta^2\big)=0\ .
\eqno(4)
$$
Subject to the convention that the positive $\phi$ domain  should be given
by positive $z$, this equation has a unique asymptotically vacuum solution,
which is given by
$$
\Phi=\eta \phi_{_{(0)}}\, \hskip 1.6 cm \phi_{_{(0)}}={\rm tanh}
\{(\eta\sqrt{2\lambda}\, z)\} \ .
\eqno(5)
$$
By substituting this in (1) and integrating over $z$ one obtains the the
constant effective action per unit measure of the worldsheet that is taken
as the basis of the (Dirac type) thin membrane model that is generally
expected to provide a good macroscopic description of the dynamical
behaviour of the wall under conditions such that that the relevant dynamical
lengthscales $L$ are all very large compared with the dimension
$$
\ell={1\over\eta\sqrt{2\lambda}}
\eqno(6)
$$
that characterises the thickness of the wall.

The question motivating the present work is how to include the corrections
to the thin membrane model that one would expect to be needed when the
relevant dimensionless curvature magnitude
$$
\epsilon={\ell\over L}
\eqno(7)
$$
is
still small, but not entirely negligible, as it must be for the simple
membrane approximation to be valid.

Starting off in the same way as in an earlier analysis{[8]} (which was more
general the present one in so much as it included allowance for weak self
gravitation) what we want do is to consider configurations obtained by
perturbing the standard solution in such a way  the coordinates parallel to
the wall are no no longer exactly but but only approximately ignorable. In
other words
$$
\ell {\partial\Phi\over\partial x^{_{\rm D}}}={\cal O}\big(1\big)\ ,
\hskip 1.6 cm \ell {\partial \Phi\over\partial x^i}={\cal O}
\big(\epsilon\big)\ ,
\eqno(8)
$$
in the limit of large values of the lengthscale $L=\ell/\epsilon$
characterising variations in directions parallel to the worldsheet of the
domain wall.

In order to proceed with the calculation, we  split quantities into their
components perpendicular and parallel to the defect worldsheet, $\Sigma$
say, where $\Phi$ vanishes in the middle of the wall. This is
done formally by utilising a Gauss-Codazzi formalism, the details of which
were developed in the earlier analysis{[8]} and are paraphrased here.

We take $n^\mu$ to be a unit geodesic normal vector field to $\Sigma$, and we
generalise the coordinate $z$ by defining it to be the proper length along the
integral curves of $n^\mu$. Each constant $z$ surface then has unit normal
$n_\mu$, fundamental tensor $h_{\mu\nu}$ (the background projection
of the intrinsic metric), and extrinsic curvature $K_{\mu\nu}$
defined by
$$
h_{\mu\nu} = g_{\mu\nu}-n_\mu n_\nu \ , \hskip 16mm
K_{\mu\nu} = -h^\rho_\mu \nabla_\rho n_\nu .
\eqno(9)
$$
Using the Gauss Codazzi formalism, the equations of motion for the wall can be
written in ``{\sevenrm D}+1'' fashion
$$
\eqalignno{
{\cal L}_n h_{\mu\nu} &= 2K_{\mu\nu} \ , & (10a) \cr
{\cal L}_n K_{\mu\nu} &= K_{\mu\rho}K^\rho_\nu \ , & (10b) \cr
{\cal L}_n {\cal L}_n \Phi + K {\cal L}_n \Phi &+ D_iD^i \Phi - 4\lambda \Phi
(\Phi^2-\eta^2) =0 \ , & (10c) \cr
}
$$
where $\sigma^i$ are taken to be coordinates on the wall, $D_i$ is the
derivative operator for the wall hypersurface, and \L$_n$ is the Lie derivative
along the vector field $n^\mu$.

\bigskip\noindent
{\bf 3. The approximation scheme.}
\medskip

The foregoing system is a complete {\it exact} set of equations for the
geometry and fields of the model,  which we now intend to analyze along
the lines described above. This means that after scaling out the dimensional
dependence on wall width and curvature, we shall make a power series
expansion of the physical quantities in terms of $\epsilon$, the ratio of
the wall width to its radius of curvature. We therefore start by setting
$$
u={z\over\ell}\ , \hskip 5mm
\Phi=\eta\phi\ ,\hskip 5mm  K_{\mu\nu}={1\over L}\kappa_{\mu\nu}\ .
\eqno(11)
$$
In terms of these new variables we have \L$_n = \ell^{-1}\partial/\partial
u$, and hence, using the abbreviation
$$
^\prime\equiv{\partial\over\partial u}\ ,
\eqno(12)
$$
we obtain
$$
\eqalignno{
h_{\mu\nu}' &= 2 \epsilon \kappa_{\mu\nu} \ , & (13a) \cr
\kappa_{\mu\nu}' &= \epsilon \kappa_{\mu\rho} \kappa^\rho_\nu \ ,
& (13b) \cr
\phi'' &- 2 \phi(\phi^2-1) +
\epsilon\kappa\phi' + \epsilon^2 D_iD^i \phi = 0 \ , &
(13c) \cr
}
$$
which is the starting point for a rigorous expansion in powers of $\epsilon$.

It is worth digressing at this point to address a misconception that has
arisen as to the interpretation of $u$ in the zero thickness limit.
Formally, ``setting $\epsilon=0$'' is interpretable as either letting the
wall thickness  vanish, or letting the wall become flat. It has been suggested
that it is incorrect to expand quantities in $\epsilon$ when $\epsilon\to0$
corresponds to the former limit, since in this limit fields become
discontinuous{[13]}. However, in the limit $\ell \to 0$, the coordinate $u$,
whilst having an infinite range, corresponds to an infinitesimal physical
range, that range being $(0^-,0^+)$ in $z$-space. Thus the coordinate $u$
takes the step function in $z$-space and `blows it up' to give a continuous
interpolation between the vacua on either side of the infinitesimally thin
wall. Thus this limit is singular only in the literary, rather than the
mathematical, sense!

We now proceed by expressing the rescaled quantities as power series in
$\epsilon$ (with coefficients that are functions of the coordinates
$\{\sigma^i,u\}$) in the form
$$
\eqalign{
\phi&=\phi_{_{(0)}}+\epsilon\, \phi_{_{(1)}}+{\epsilon^2\over 2}
\phi_{_{(2)}}+{\cal O}\{\epsilon^3\}\ ,\cr
h_{\mu\nu} &= h_{_{(0)}\mu\nu} + \epsilon h_{_{(1)}\mu\nu}+ {\epsilon^2\over2}
h_{_{_(2)}\mu\nu} + {\cal O}\{\epsilon^3\} \ , \cr
}
 \eqno (14)
$$
and
$$
\kappa_{\mu\nu}=\kappa_{_{(1)}\mu\nu}+{\epsilon\over 2}
\kappa_{_{(2)}\mu\nu}+{\epsilon^2\over 6}\kappa_{_{(3)}\mu\nu}+{\cal O}
\{\epsilon^4\}\ .
\eqno(15)
$$
Substituting such a power expansion into (13) gives a sequence of equations
obtained by setting the coefficients of successive powers of $\epsilon$ to
zero.

To zeroth order, the geometry is independent of $u$, and the field equation
reduces to (4), which is automatically satisfied by using the expression (5)
for $\phi_{_{(0)}}$, which, in terms of the rescaled coordinate $u$, is
simply
$$
\phi_{_{(0)}}={\rm tanh}\, u\ .
\eqno(16)
$$

After the lowest order requirement (4), the next (the last that will be
needed here) in the sequence of requirements obtained from (13c) is the one
governing the first order field $\phi_{_{(0}}$, which satisfies the
dynamical equation
$$
\phi_{_{(1)}}^{\ \prime\prime}-2\big(3\phi_{_{(0)}}^{\ 2}-1\big)
\phi_{_{(1)}} =-\kappa_{_{(1)}}\phi_{_{(0)}}^{\ \prime}\ .
\eqno(17)
$$
The driving term on the right of this linearised perturbation equation can
be seen to be proportional to the lowest order coefficient in the expansion
for the extrinsic curvature scalar.

\bigskip\noindent
{\bf 4. The question of field regularity on the defect locus.}
\medskip

Up to this point the present analysis agrees completely with that of the
previous work{[8 -- 10]}, which went on from here to make the {\it crucial
observation} that unless the scalar curvature coefficient $\kappa_{_{(1)}}$
vanishes on the wall, the equation (17) has no solution that is regular and
bounded over the whole range from $u=-\infty$ to $u=+\infty$. It can be
deduced from this that freely moving domain walls satisfying the field
equations must obey the condition
$$
\kappa_{_{(1)}}=0\ .
\eqno(18)
$$
This is exactly what is required for consistency with the thin (Dirac type)
membrane treatment of the dynamics in the extreme limit when $L/\ell$
is very large, for which the dynamic equations are well known to consist
just of the ``harmonicity'' condition to the effect that the trace of the
membrane curvature scalar $K$ should vanish.

It is at the next stage of the work that discord arises. The ultimate motive
for the present work, as indeed for previous work,  is the derivation of
higher order corrections to the simple Dirac membrane approximation. The
obviously natural and generally agreed strategy for doing this is to try to
apply the same kind of procedure that was used in the zero order membrane
treatment whereby the spacetime action integral
$$
{\cal I}=\int {\hat{\cal L}}\,\sqrt{-{\rm det}\, g}\, d^{\rm D +1}x
\eqno(19)
$$
is expressed in the form
$$
{\cal I}=\int{\cal L}\,\sqrt{-{\rm det}\, h\{\sigma\} }\, d^{\rm D}\sigma\ ,
\eqno(20)
$$
in which the off-worldsheet degrees of field freedom are eliminated from the
worldsheet hypersurface Lagrangian density ${\cal L}$ which is to be
obtained by integrating the ordinary spacetime Lagrangian density
${\hat{\cal L}}$ over the remaining dimension parametrised by $z$ that is
suppressed in (20) after fixing the off-wall values of the field variables
by the requirement that the off-wall field equations should be satisfied to
the required degree of accuracy.

Where this paper departs from previous work{[8 -- 12]} is in the use made
of the {\it crucial observation} cited above:
on the basis of the supposition that the solution of (17)
should  be regular and bounded over $u\in $ \real, it was
argued previously that (18) should indeed be satisfied, i.e.  that
$\kappa_{_{(1)}}$ {\it must} vanish. This is, in essence, a requirement that
the field equations should be satisfied not just {\it off} the perturbed
worldsheet but even {\it on} it. If we were already trying to solve for the
motion of the wall, this would be an eminently reasonable, and indeed
necessary, step to take, but we have not yet got to that stage. The aim of
the game at this stage is to try to find an effective wall action that will
be varied later on to get the equations of motion of the wall location. We
must therefore be careful that we solve, or eliminate, only those degrees of
freedom which are external to the wall, maintaining the fully unrestrained
``off-shell'' character of those virtual modes corresponding to the degrees
of freedom of the wall itself. The more severe requirement postulated in the
previous work{[8 -- 10]} is interpretable as demanding that the worldsheet
should satisfy the relevant dynamical equation -- namely (18) in the present
instance -- which is clearly not consistent with the requirement that the
``off shell" world sheet configuration in the action (20) should be freely
variable. The premature imposition of the dynamical condition (18) resulted
in the unjustified suppression{[8 -- 12]} of a potentially important
contribution to the action that needs to be evaluated. In order to avoid
premature imposition of the dynamical equation (18) when evaluating the
action one {\it must not} try to satisfy the first order field equation (17)
continuously over the whole range extending through the defect locus
$\Sigma$ itself where $\Phi$ vanishes, but only in the separate domains
outside this locus.

\bigskip\noindent
{\bf 5. Evaluation of the linearised solution and the corresponding action.}
\medskip

It follows from the preceding considerations that the
 appropriate procedure is just to require that the field equation be
satisfied separately in the positive $\Phi$ domain $0<u<\infty$ and in the
negative $\Phi$ domain $-\infty<u<0$, i.e., off the defect locus.
The boundary conditions localising the defect $\Phi=0$ at the middle
of the wall where $u=0$ and imposing a vacuum state at infinity
are expressible formally as
$$
\eqalignno{
\lim_{u\to 0^\pm} \Phi = 0 \hskip 1 cm &\Rightarrow\hskip 1cm
\phi_{_{(0)}} \rightarrow 0, \ \phi_{_{(1)}}\rightarrow 0,  ...\
{\rm as }\ u\to 0^\pm, & (21a)\cr
\lim_{u\to \pm\infty} \Phi = \pm \eta \hskip 1cm &\Rightarrow\hskip 1 cm
\phi_{_{(0)}}\rightarrow 1, \ \phi_{_{(1)}}\rightarrow 0,  ...\ {\rm as }\
u\to\pm\infty .&(21b)\cr }
$$

Subject to the foregoing requirements,
the linearised field equation (17) is uniquely soluble. The required
solution is given, without any ambiguity at all, by
$$
\phi_{_{(1)}}=\kappa_{_{(1)}}\, f \ ,
\eqno(22)
$$
where the dimensionless function $f$ of $u$ has the explicit analytic form
$$
f\{u\}=\pm{1\over 2}\,{\rm tanh}\{u\}-{1\over 2}+\Big({2\over 3}\pm{u\over
2}\Big) {1\over{\rm cosh}^2\{u\} }-{1\over 6}\,{\rm exp}\{\mp 2u\} \
,\eqno(23)
$$
in which the upper and lower sign choices apply respectively to the positive
and negative domains, so that $f$ is even under reflection, i.e.
$f\{u\}=f\{-u\}$. At the origin $u=0$ separating the two domains this
function is constructed so as to vanish, $f\{0\}=0$, but its gradient there
has a non vanishing limit, $\big( df/ du\big)\big |_0=\pm {4\over3}$ so that it
has a
discontinuity across the wall given by $\big[df/du\big]^+_-=8/3$.
\midinsert \hskip 7 truemm \epsfbox{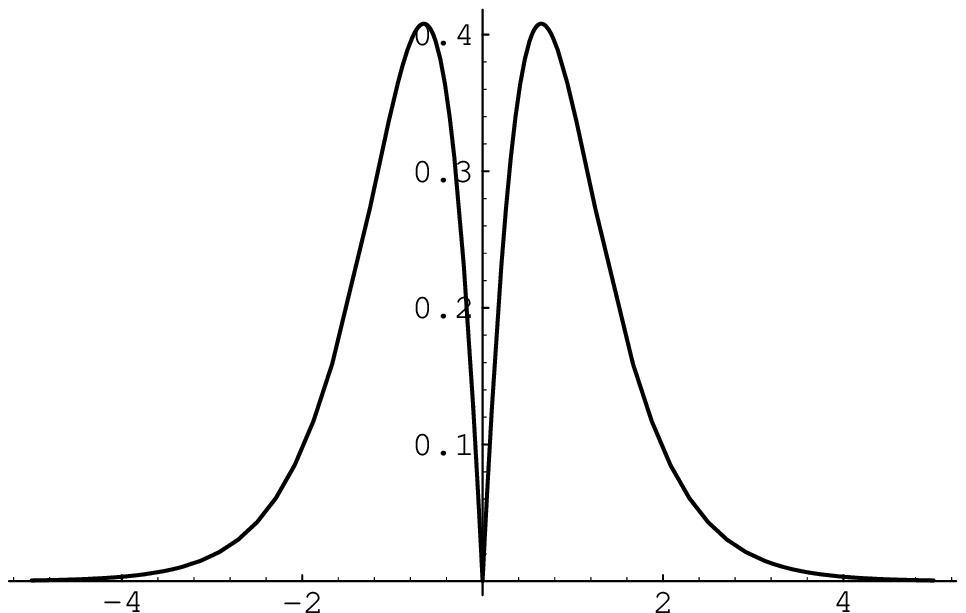} \medskip \hskip 2 truecm
\vbox{ \hsize=11.5 truecm
\noindent {\cat FIGURE (1): Numerical evaluation of the function $f$.}}
\endinsert

In terms of the dimensionless function $f$, the solution for $\Phi$ itself
(which is thus continuous but not continuously differentiable across the
wall surface  $z=0$) will be given to the required order, with the
dimensional parameters restored, by
$$
\Phi=\eta\,{\rm tanh}\,\big\{ {z\over\ell}\big\}+\eta\ell K\,f\big\{
{z\over\ell}\big\}+{\cal O}\{\epsilon^2\} \ .
\eqno(24)
$$
\midinsert \hskip 7 truemm \epsfbox{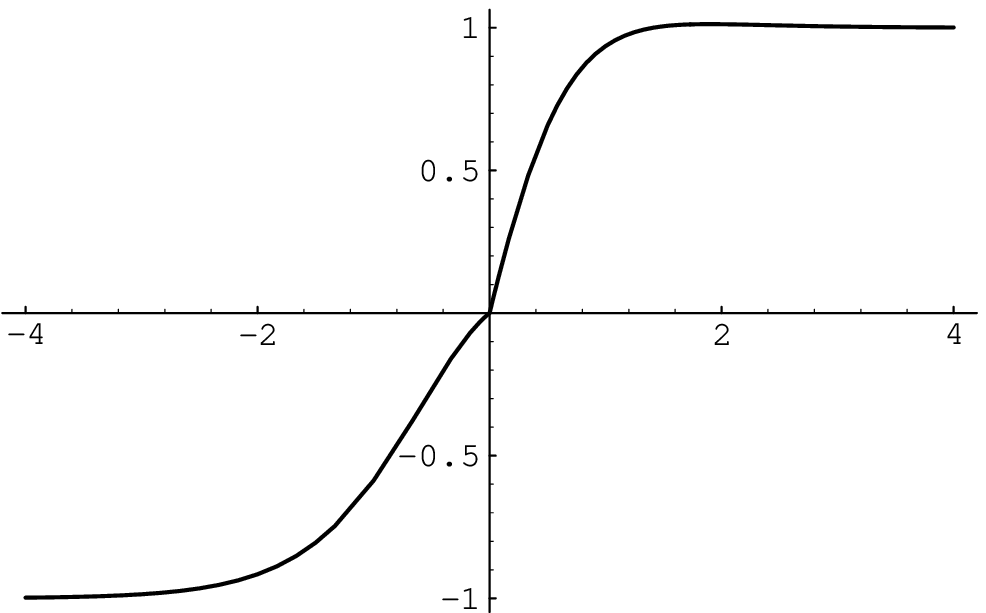} \medskip \hskip 2 truecm
\openup-2\jot \vbox{ \hsize=11.5 truecm \noindent {\cat FIGURE (2):
Approximate evaluation of the dimensionless field $\phi$  for the
(rather large) perturbation amplitude $\ell K=0.5$.}}  \openup 2\jot \endinsert

In terms of the solution (24) it is now straightforward to evaluate the
corresponding expression for the effective domain wall surface Lagrangian
${\cal L}$ in (20) which will be given by
$$
{\cal L} \equiv {\cal L}\{\sigma\}=\int{\hat{\cal L}}\, J\, dz \ ,
\eqno(25)
$$
where ${\hat{\cal L}}$ is the original Lagrangian density function (1) as
evaluated for the solution (24), and $J$ is the relevant Jacobean factor
which is given by
$$
J={\sqrt{-{\rm det}\, g}\hskip 6mm\over \sqrt{-{\rm det}\,
h}\big |_{u=0}} \ .
\eqno(26)
$$

Since the first order contribution to \L~will vanish by the zeroth order field
equations, it is necessary to work out (25) to second order to get the
lowest non-trivial corrections to the simple Dirac membrane treatment. To
this degree of accuracy, the geometry is readily calculated from (13) as
$$
h_{_{(0)}\mu\nu} =h_{_{(0)}\mu\nu} (\sigma)\ ; \hskip 5mm h_{_{(1)}\mu\nu} =
2u\kappa_{_{(1)}\mu\nu}  \ ; \hskip 5mm h_{_{(0)}\mu\nu}  = 2u^2
\kappa_{_{(1)}\rho\nu} \kappa_{_{(1)}\mu }^\rho \ ,
\eqno (27a)
$$
$$
\kappa_{_{(1)}\mu\nu}= \kappa_{_{(1)}\mu\nu}\{\sigma\}\ ; \hskip 10mm
\kappa_{_{(2)}\mu\nu}=2u\kappa_{_{(1)}\mu\rho}\kappa_{_{(1)} \nu}^\rho \ ,
\eqno (27b)
$$
and hence the Jacobean (26) is obtained via
$$
\eqalign{
\sqrt{-g} &= \sqrt{-g}|_{u=0} + \epsilon u (\sqrt{-g})'|_{u=0} + \epsilon^2
(\sqrt{-g})''|_{u=0} + ...\cr
&= \sqrt{- h}|_{u=0} \left [ 1 + \epsilon u \kappa_{_{(1)}} + {\epsilon^2 u^2
\over 2} (\kappa_{_{(1)}}^2 - \kappa_{_{(1)}\mu\nu}
\kappa_{_{(1)}}^{\ \mu\nu}) + ...\right ]
\cr
}
\eqno (28)
$$
as
$$
J = 1 + \epsilon u \kappa_{_{(1)}} + {\epsilon^2 u^2
\over 2} (\kappa_{_{(1)}}^2 - \kappa_{_{(1)}\mu\nu}
\kappa_{_{(1)}}^{\ \mu\nu}) + ... \ .
\eqno (29)
$$

Since $\phi_{_{(0)}}$ depends only on $u$ and $\phi_{_{(1)}}$ depends on
the other coordinates $\sigma^i$ only through $\kappa_{_{(1)}}$  it follows
that
we have $\partial\phi_{_{(0)}}/\partial x^i=0$ and
$\partial\phi_{_{(1)}}/\partial x^i={\cal O}\{\epsilon\}$, and hence that
$\partial\phi/\partial x^i={\cal O}\{\epsilon^2\}$. This implies that up to
(and even beyond) the required degree of accuracy  the Lagrangian density
${\hat{\cal L}}$ will be expressible simply as
$$
{\hat{\cal L}}=-\lambda\eta^4\Big(\phi^{\prime\, 2}+\big(\phi^2-1\big)^2\Big)+
{\cal O}\{\epsilon^4\} \ .
\eqno(30)
$$
We thus obtain
$$ {\hat{\cal L}}={\hat{\cal L}}_{_{(0)}}+\epsilon\, {\hat{\cal
L}}_{_{(1)}}+{\epsilon^2\over 2} {\hat{\cal L}}_{_{(2)}} + {\cal
O}\{\epsilon^3\}\ ,
\eqno (31)
$$
with
$$
{\hat{\cal L}}_{_{(0)}} = -\lambda\eta^4\Big(\phi_{_{(0)}}^{\prime\ 2}+\big(
\phi_{_{(0)}}^{\ \ 2}-1\big)^2\Big)= -2\lambda\eta^4\phi_{_{(0)}}^{\prime\ 2}
\ ,
\eqno(32a)
$$
$$
{\hat{\cal L}}_{_{(1)}} = -2\lambda\eta^4\Big(\phi_{_{(0)}}^{\prime}
\phi_{_{(1)}}^{\prime}+2\big(\phi_{_{(0)}}^{\ \ 2}-1\big)\phi_{_{(0)}}
\phi_{_{(1)}}\Big) = -2 \lambda \eta^4 ( \phi_{_{(0)}}'\phi_{_{(1)}})'\ ,
\eqno(32b)
$$
and
$$
\eqalign{
{\hat{\cal L}}_{_{(2)}} &= -\lambda\eta^4\Big(2\phi_{_{(0)}}^{\prime}
\phi_{_{(2)}}^{\prime}+ \phi_{_{(1)}}^{\prime\ 2}
+4\big(\phi_{_{(0)}}^{\ \ 2}-1\big)\phi_{_{(0)}}\phi_{_{(2)}}
+2\big(3\phi_{_{(0)}}^{\ \ 2}-1\big)\phi_{_{(1)}}^{\ \ 2}\Big)\cr
&= - \lambda\eta^4 \Big ( 2(\phi_{_{(0)}}'\phi_{_{(2)}})' +
(\phi_{_{(1)}}'\phi_{_{(1)}})' + \kappa_{_{(1)}} \phi_{_{(1)}}
\phi_{_{(0)}}' \Bigr ) \ , \cr
}
\eqno(32c)
$$
using the field equations (4) and (17).

Using these expressions  to simplify the corresponding
expansion
$$
{\hat{\cal L}} J = {\hat{\cal L}}_{_{(0)}} \Big (
1 + \epsilon J_{_{(1)}} + {\epsilon^2\over 2}  J_{_{(2)}} \Big )
+ {\hat{\cal L}}_{_{(1)}} \Big ( 1 + \epsilon J_{_{(1)}} \Big )
+ {\epsilon^2\over 2} {\hat{\cal L}}_{_{(2)}} + {\cal O}\{\epsilon^3\}\ ,
\eqno(33)
$$
the required integrand ${\hat{\cal L}} J$ in (25) is found to be
expressible to  the required degree of accuracy by
$$
{{\hat{\cal L}} J \over \lambda\eta^4} = -2\phi_{_{(0)}}^{\prime\ 2}
\Big ( 1+ \epsilon J_{_{(1)}} + {\epsilon^2\over 2} J_{_{(2)}} \Big )
+ \epsilon^2 \kappa_{_{(1)}} \phi_{_{(0)}}^{\prime} \phi_{_{(1)}}
$$
$$
-\epsilon \Big ( \phi_{_{(0)}}^{\prime} \big ( 2\phi_{_{(1)}}
+\epsilon\phi_{_{(2)}} \big ) + {\epsilon\over 2} \big (
\phi_{_{(1)}}^{\prime} + 2 \kappa_{_{(1)}} \phi_{_{(0)}} u \big )
\phi_{_{(1)}} \Big ) ^\prime + {\cal O}\{\epsilon^3\}\
\eqno (34)
$$
in each of the separate domains $-\infty<u<0$ and $0<u<\infty$.

The integral (25) will be expressible as the sum of contributions from
each of the two separate domains in the form
$$
{\cal L}=\int_{_\infty}^0{\hat{\cal L}} J\, du +
\int_0^{\infty}{\hat{\cal L}} J\, du \ .
\eqno(35)
$$
The condition that $\phi$ and hence also the separate expansion
coefficients  $\phi_{_{(0)}}$, $\phi_{_{(1)}}$ and $\phi_{_{(2)}}$ should
vanish at the domain boundary $u=0$, together with the outer limit
condition $\phi-\phi_{_{(0)}}\rightarrow 0$, which implies $\phi_{_{(1)}}
\rightarrow 0$ and $\phi_{_{(2)}}\rightarrow 0$, as $u\rightarrow
\pm\infty$, implies that there is no contribution from the total derivative
in (34). It can also be seen that the first order contribution of the integrand
is an odd function of $u$ and thus that it cancels out  between the two terms
in
(35), so that the final result is obtained in the expected form
$$
{\cal L}= {\cal L}_{_{(0)}}+{\epsilon^2\over 2}{\cal L}_{_{(2)}}
+{\cal O}\{\epsilon^3\} \ ,
\eqno (36)
$$
with
$$
{\cal L}_{_{(0)}}=-{2\over\ell}\eta^2 I_{\rm I} \ ,
\eqno(37)
$$
and
$$
{\cal L}_{_{(2)}}={-2\over\ell}\eta^2\big (\kappa_{_{(1)}}^{\, 2}-
 \kappa_{_{(1)}\mu\nu}\kappa_{_{(1)}}^{\ \mu\nu} \big) I_{\rm I\!I}
+{2\over\ell}\eta^2\kappa_{_{(1)}}^{\, 2}I_{\rm I\!I\!I} \ ,
\eqno(38)
$$
where the dimensionless constant coefficients are given as the integrals
$$
\eqalignno{
I_{\rm I} &= \int_{-\infty}^\infty \phi_{_{(0)}}^{\prime\ 2}\, du
= {4\over3} \ , & (39a) \cr
I_{\rm I\!I} &= \int_{-\infty}^\infty \phi_{_{(0)}}^{\prime\ 2} u^2\, du
= {\pi^2-6\over 9}\ , & (39b) \cr
I_{\rm I\!I\!I} &= \int_{-\infty}^\infty \phi_{_{(0)}}^{\prime}\, f \,du
= {8\over 9} \ . & (39c) \cr
}
$$
The difference between the present calculation and its predecessors{[8 -- 12]}
is the inclusion here of the extra term proportional to $I_{\rm I\!I\!I}$ in
(38).

\bigskip\noindent
{\bf 6. The canonically truncated model.}
\medskip

The outcome of the preceding  calculation is that the second order
effective action obtained for the wall from (36) by truncating the
uncalculated higher order correction ${\cal O}\{\epsilon^3\}$ will be
expressible explicitly, with the dimensional factors restored, as
$$
{\cal L}=-{8\over3\ell}\eta^2 \left [ 1 + {\rm C}_{\rm I}
R + {\rm C}_{\rm I\!I} K^2 \right ] \  ,
\eqno(40)
$$
where $R$ is the 3-dimensional Ricci scalar of the internal
metric $h_{ij}$ of the wall, which is given by the well known Gauss
formula
$$
R= K^{2} - K^\mu_\nu K^\nu_\mu \ ,
\eqno(41)
$$
while the coefficients are constants, of the order of the square of
the wall thickness $\ell$, which are given exactly by
$$
{\rm C}_{\rm I}={ I_{\rm I\!I}\over I_{\rm I}  }{\ell^2\over 2}
={\pi^2-6\over 24}\ell^2\ ,\hskip 1 cm {\rm C}_{\rm I\!I}
=-{ I_{\rm I\!I\!I}\over I_{\rm I}  }{\ell^2\over 2} =-{1\over 3}\ell^2 \ .
\eqno(42)$$

Using the formula (A10) obtained in earlier work{[9]} (after rectification
of a transcription error interchanging the parameters $\beta$ and $\Delta$
that were then to be identified) or more rapidly by direct substitution of
the expressions $K_{\mu\nu}^{\ \ \,\rho}=K_{\mu\nu}n^\rho$ and $K^\rho=K
n^\rho$ (for the second fundamental tensor and its contraction) in the
general (dimensionally unrestricted) formulae that have been derived more
recently{[15]}, the equation of motion that ultimately results from the
Lagrangian (40) is found to be given by
$$
K={\rm C}_{\rm I}\big(3K K^\mu_\nu K^\nu_\mu -K^3-
2K^\mu_\nu K^\nu_\rho K^\rho_\mu\big)+{\rm C}_{\rm I\!I}\big(2 K
K^\mu_\nu K^\nu_\mu-K^3+ 2\bo K\big)\ ,
\eqno(43)
$$
(where $\bo$ denotes the worldsheet Dalembertian) in which the final
bracket, with coefficient  ${\rm C}_{\rm I\!I}$, groups the contributions
that unjustifiably left out in the previous work{[8 -- 12]}. It is to be
remarked that one is free to work with units that adjust the numerical
value of the length scale $\ell$ in order to set either of the magnitudes
(though not the signs) of the coefficients ${\rm C}_{\rm I}$ and ${\rm
C}_{\rm I\!I}$ to any chosen value such as unity: thus apart from the signs
(which, as discussed below, are of crucial importance) all that matters
qualitatively is their magnitude ratio, $c$ say, which is given by
$$
c=-{{\rm C}_{\rm I\!I}\over{\rm C}_{\rm I}}=
{I_{\rm I\!I\!I}\over I_{\rm I\!I}}={8\over \pi^2-6}\simeq 2 \ .
\eqno(44)$$

It is to be remarked that (unlike what can be seen to occur in the string
case{[15]} due to the divergence property of its Ricci scalar)
the exact satisfaction of the lowest order dynamical equation, namely
$K=0$, is {\it not} by itself sufficient to ensure satisfaction of the
corresponding higher order system (43). The simple harmonicity condition
$K=0$ can however be seen to be sufficient in the restricted case of
a {\it static} configuration in ordinary flat spacetime (with D=3) since
in these circumstances it automatically entails the cubic
order condition $K^\mu_\nu K^\nu_\rho K^\rho_\mu=0$ as well, which is
evidently enough.

\bigskip\noindent
{\bf 7. Implications.}
\medskip

The lowest order contribution, $-8\eta^2 /3\ell$, to the Lagrangian (40) is
the constant that by itself gives the simple Dirac membrane action. The next
term, proportional to the worldsheet Ricci scalar $R$ with coefficient
$-(\pi^2-6)\eta^2 \ell/9$, is the purely ``geometric" contribution  whose
derivation is described in the previous work{[8 -- 10]} that was cited
above. However that work overlooked the final ``deformation" term,
proportional to $K^2$ with coefficient $8\eta^2\ell/9 $, which arises from
the first order correction term in (24) when this expression is substituted
in (1) prior to performance of the integration over the off worldsheet
dimension parametrised by $z$. In the more familiar example of buckling in a
bent elastic rod, the deformation correction reduces the bending energy
arising from the rigidity of the solid material involved and is therefore
appropriately describable as an ``antirigidity" effect. It is therefore
reasonable also to  describe the negativity of the coefficient ${\rm C}_{\rm
I\!I}$ for the analogous deformation term in the present example as an
``antirigidity" effect.

The idea implicit in the above terminology is that the contribution to the
energy in a static configuration (that is not necessarily a solution) should
be positive in the case of a ``rigidity" term and negative in the case of an
``antirigidity" term. However one should be aware that the notion of
``rigidity" (whose introduction in the present context is attributable to
Polyakov${[14]}$) is potentially misleading since one can conceive
alternative defining conventions in terms of criteria for stable
equilibrium, for which however the alternative term ``stiffness" is perhaps
more appropriate. A systematic study${[15]}$ of the effect of conceivable
quadratic curvature corrections for closed maximally symmetric p-brane
configurations -- meaning a circles in the case of a string with p=1,
spheres in the case of a membrane with p=2, and so on in hypothetical higher
dimensional cases -- shows that in the case of strings the criterion of
positive ``rigidity" according to the defining convention postulated above
agrees with the condition for the existence of static ring solutions,
i.e.  it is positive ``rigidity" that provides ``stiffness".  On the
other hand in higher dimensional cases with p $>$ 2 it is ``antirigidity" as
defined above that is required for the existence of static equilibrium: the
negativity of ${\rm C}_{\rm I\!I}$ in the theory considered here
is thus interpretable as making a higher dimensional wall ``stiff" in the
sense of allowing it to avoid collapse in a hyperspherical configuration.
However this notion of ``stiffness'' loses its meaning in the critical
intermediate case, with p=2, that applies to walls in ordinary 4-dimensional
spacetime, for which spherical equilibrium will {\it always be impossible},
regardless of whether the sign of the coefficient ${\rm C}_{\rm I\!I}$
is positive, which would correspond to ``rigidity", or negative as in the
specific ``antirigid" wall model considered here.

 The question of existence of static solutions leads on to the question of
their stability. Although the model characterised by (40) has no maximally
symmetric static solution that is closed within a 4-dimensional spacetime
background, it is evident that there will always be one that is open,
namely the simple plane wall solution. It is also evident that at least
locally there will be many other static, though less highly symmetric
solutions, whose stability can be tested by linear perturbation theory.
Although it may have some effect on their propagation speeds, the extra
``antirigidity" term evidently cannot destabilize any of the large $L$ (i.e.
low frequency, long wavelength) modes to which the validity of our
derivation of the model (40) is restricted. However the negativity of ${\rm
C}_{\rm I\! I}$ will engender instability in  modes whose characteristic
curvature scale $L$ is small enough to be comparable with the wall width
$\ell$. This instability in the model characterised by (40) and (42) does
not mean that the domain wall is actually unstable: it merely means that
such rapidly varying modes cannot be treated adequately without allowance
for the higher order terms ${\cal O}\{\epsilon^3\}$ that were thrown away in
the truncation that was made in going from (36) to (40). This feature is
a serious drawback from the point of view of the use of (40)
in conjunction with (42) in practice:
it implies the need, in numerical computations, to incorporate some
artificial mechanism for damping out the unphysical short timescale
instabilities that would otherwise occur.

This caveat to the effect that the canonically
truncated model of the previous section should not be
taken too literally but used with caution, provides the motivation
for seeking a more practically convenient alternative.
A reasonable way of getting round the difficulty in the practical
calculation of curvature corrections to domain wall dynamics in the
long wavelength limit is to take advantage of the reassuring observation
that, whereas it only has to satisfy $\ell K={\cal O}\{\epsilon\}$
``off shell", this dimensionless combination must satisfy the more
severe requirement
$$\ell K={\cal O}\{\epsilon^3\} \eqno(7.1)$$
for any configuration that is actually a solution
of the dynamical equation (43). The corresponding reduced curvature
scalar must therefore satisfy $\kappa={\cal O}\{\epsilon^2\}\ $,
the latter being expressible, by (15), as the vanishing not only of
$\kappa_{_{(1)}}$ but even of $\kappa_{_{(2)}}$, which is more than
enough to ensure that the litigious regularity condition (18) (that was
imposed prematurely, before the variation, in the previous work {[8 -- 10]})
will after all be satisfied ``on shell" as one would expect. It follows that
whether it be obtained from the truncated Lagrangian (40) or from the
original expansion (36) a solution of the dynamical equations will be
characterised up to second order corrections by
$$\ell K+{\pi^2-6\over 12} \ell^3 K^\mu_\nu K^\nu_\rho K^\rho_\mu={\cal O}
\{\epsilon^4\} \eqno(7.2)$$
This is evidently the same as would be obtained by taking the deformation
coefficient ${\rm C}_{\rm I\!I}$ to vanish, i.e. setting $c=0$ as was done
in the previous work{[8 -- 12]} instead of using the value $c\simeq 2$
derived by the logically consistent procedure used above.

The conclusion is that although, strictly speaking, the internal mechanics
of the wall is really characterised by the ``antirigidity" property
represented by the well defined negative value of ${\rm C}_{\rm I\!I}$ as
given by (43), nevertheless this effect does not influence the dynamics to
the order of accuracy under consideration. It is therefore quite permissible
to use the simpler and better behaved ``zero rigidity" model specified by
setting ${\rm C}_{\rm I\!I}=0$ in (40) as advocated in the previous work{[8
-- 12]}. There is however nothing obligatory about this option: it would
also be permissible (for example if it were thought helpful for numerical
computations) to use an overstabilised model characterised by a positive
value of ${\rm C}_{\rm I\!I}$ provided it did not exceed the order of
magnitude limitation $\vert {\rm C}_{\rm I\!I}\vert \lta \ell^2$.

\bigskip
\noindent{\bf Acknowledgements.}
\medskip

The authors wish to thank C. Barrab\`es, D. Garfinkle, D. Haws, T. Kibble,
K. Maeda, X. Martin, P. Peter, M. Sakellariadou, J. Stewart, and N. Turok
for pertinent conversations on various past and recent occasions.

\vfill\eject

\bigskip\parindent= 0 cm
\noindent{\bf References.}

\medskip

\nref T. Vachaspati and A. Achucarro, \prd 44 3067 1991.

\nref N. Turok, \prl 63 2625 1989.

\nref B. Carter, {\it Ann. New York Acad. Sci.} {\bf 647}, 758 (1992).

\nref D. Forster, \npb 81 84 1974.

\nref K. Maeda, N. Turok, \plb 3 376 1988.

\nref R. Gregory, \plb 199 206 1988.

\nref V. Silveira, M.D. Maia, \pla 174 280 1993.

\nref D. Garfinkle, R. Gregory, \prd 41 1889 1990.

\nref R. Gregory, D. Haws, D. Garfinkle, \prd 42 343 1990.

\nref R. Gregory, \prd 43 520 1991.

\nref P.S. Letelier, \prd 41 1333 1990.

\nref C. Barrab\`es, B. Boisseau, M. Sakellariadou, \prd 49 2734 1994.

\nref A. K. Raychaudhuri, G. Mukherjee, \prl 59 1504 1987.

\nref A. Polyakov, \npb 268 406 1986.

\nref B. Carter, \book Equations of motion of a stiff geodynamic string or
higher brane. [[1994 Meudon preprint, to appear in {\it Classical and
Quantum Gravity} {\bf 11}.]]

\end